\documentclass[preprint,aps,superscriptaddress]{revtex4}

\usepackage[dvipdfmx]{graphicx}
\usepackage{dcolumn}
\usepackage{bm}
\usepackage{color}

\begin{document}

\preprint{APS/123-QED}

\title{V 3$d$ charge and orbital states in V$_2$OPO$_4$ probed by x-ray absorption spectroscopy}

\author{Kota~Murota}
\affiliation{Department of Applied Physics, Waseda University, Shinjuku, Tokyo 169-8555, Japan}
\author{Elise~Pachoud}
\affiliation{Centre for Science at Extreme Conditions and School of Chemistry, University of Edinburgh, Edinburgh EH9 3FD, United Kingdom}
\author{J.~Paul~Attfield}
\affiliation{Centre for Science at Extreme Conditions and School of Chemistry, University of Edinburgh, Edinburgh EH9 3FD, United Kingdom}
\author{Robert~Glaum}
\affiliation{Institut f{\rm $\ddot{u}$}r Anorganische Chemie, Universit{\rm $\ddot{a}$}t Bonn, D-53012 Bonn, Germany}
\author{Ronny~Sutarto}
\affiliation{Canadian Light Source, Saskatoon, Saskatchewan S7N 0X4, Canada}
\author{Kou~Takubo}
\affiliation{Department of Chemistry, Tokyo Institute of Technology, Meguro, Tokyo 152-8551, Japan}
\author{Daniel~I.~Khomskii}
\affiliation{II Physikalisches Institut, Universit{\rm $\ddot{a}$}t zu K{\rm $\ddot{o}$}ln, 50937 K{\rm $\ddot{o}$}ln, Germany}
\author{Takashi~Mizokawa}
\affiliation{Department of Applied Physics, Waseda University, Shinjuku, Tokyo 169-8555, Japan}

\date{\today}

\begin{abstract}
V 3$d$ charge and orbital states in V$_2$OPO$_4$ have been investigated by means of x-ray absorption spectroscopy (XAS). The electronic structure of V$_2$OPO$_4$ is very unique in that the charge transfer between V$^{2+}$ and V$^{3+}$ in face sharing VO$_6$ chains provides negative thermal expansion as reported by Pachoud {\it et al.} [J. Am. Chem. Soc. {\bf 140}, 636 (2018).] The near edge region of O 1$s$ XAS exhibits the three features which can be assigned to transitions to O 2$p$ mixed into the unoccupied V 3$d$ $t_{2g}$ and $e_{g}$ orbitals of V$^{2+}$ and V$^{3+}$. The V 2$p$ XAS line shape can be reproduced by multiplet calculations for a mixed valence state with V$^{2+}$ and V$^{3+}$. The polarization dependence of the O 1$s$ and V 2$p$ XAS spectra indicates V 3$d$ orbital order in which $xy$ and $yz$ (or $zx$) orbitals are occupied at the V$^{3+}$ site in the face sharing chains. The occupied $xy$ orbital is essential for the antiferromagnetic coupling between the V$^{2+}$ and V$^{3+}$ sites along the chains while the occupied $yz$ (or $zx$) orbital provides the antiferromagnetic coupling between the V$^{2+}$ and V$^{3+}$ sites between the chains.
\end{abstract}

\pacs{   }
\maketitle

\newpage

\section{Introduction}
Transition-metal compounds have been attracting great interest due to their rich lattice and electronic properties which are derived from the transition-metal $d$ and ligand $p$ orbitals \cite{Imada1998, Khomskii2014}. The interplay between the lattice and electronic properties provides an exotic phenomenon known as negative thermal expansion (NTE) through rigid bond formation, magnetoelastic effect, and charge transfer effect \cite{Evans1999,Sleight1998,Takenaka2012,Chen2015}.
The charge transfer mechanism for NTE is characterized by a dramatic valence change of transition metals. For example, BiNiO$_3$ exhibits a valence transition from insulating Bi$^{3+}_{0.5}$Bi$^{5+}_{0.5}$Ni$^{2+}$O$_3$ to  metallic Bi$^{3+}$Ni$^{3+}$O$_3$ under high pressure \cite{Ishiwata2002, Azuma2007, Mizumaki2009}. La substitution for Bi suppresses the insulating state, and the moderate NTE is realized at ambient pressure in Bi$_{1-x}$La$_{x}$NiO$_3$ \cite{Ishiwata2005, Azuma2011, Oka2013}

Very recently, Pachoud {\it et al.} have reported NTE in V$_2$OPO$_4$ which is driven by charge transfer between V$^{2+}$ and V$^{3+}$ sites \cite{Pachoud2018}. V$_2$OPO$_4$ consists of face and corner sharing VO$_6$ octahedra as shown in Fig. 1. Below 605 K, the V$^{2+}$/V$^{3+}$ charge ordering is accompanied by monoclinic lattice distortion \cite{Pachoud2018}. 
The face sharing V$^{2+}$ and V$^{3+}$ sites form chains along the [110] or [1$\bar{1}$0] direction of the monoclinic lattice as indicated by the dashed lines in Fig. 1. The corner sharing V$^{3+}$ sites are connected approximately along the [001] direction. The VO$_6$ octahedra in the [1$\bar{1}$0] chains are tilted relative to those in the [110] chains. In Fig. 1, the $x$, $y$, and $z$ axes are along the V-O bonds of the V$^{3+}$ site in the [110] chain while the $x'$, $y'$, and $z'$ axes are along the V-O bonds of the V$^{3+}$ site in the [1$\bar{1}$0] chain.
The V$^{2+}$/V$^{3+}$ spins are ferrimagnetically ordered below 165 K. The V$^{2+}$ and V$^{3+}$ spins are antiferromagnetically coupled along the face sharing bond and the V$^{3+}$ spins are ferromagnetically coupled along the corner sharing bond \cite{Pachoud2018}. Above 165 K, the magnetic susceptibility shows a paramagnetic moment of 1.61 $\mu_B$ per V$_2$OPO$_4$ unit, which is greatly reduced from the ideal value for V$^{2+}$ and V$^{3+}$ spins \cite{Pachoud2018}. Above 605 K, the charge ordering disappears and all the V sites become V$^{2.5+}$ in tetragonal lattice structure. Since tetragonal phase with V$^{2.5+}$ has smaller volume than the monoclinic phase with V$^{2+}$ and V$^{3+}$, NTE is realized around 605 K \cite{Pachoud2018}. The NTE in V$_2$OPO$_4$ is seen at relatively high temperature compared to Bi$_{1-x}$La$_x$NiO$_3$. As for the mechanism of the charge transfer transition, the availability of single crystals enables to study role of V 3$d$ orbital ordering. Soft x-ray absorption spectroscopy is a powerful technique to study valence and orbital states of V. In the present work, we study V 3$d$ orbital states by means of polarization dependent x-ray absorption measurement and cluster model analysis. 

\section{Methods}
X-ray absorption spectroscopy (XAS) measurement was performed at the REIXS beamline of the Canadian Light Source \cite{Hawthorn2011}. The incident soft x-ray is linearly polarized. XAS spectra were taken in the total electron yield (TEY) mode and total fluorescence yield (TFY) mode. The single crystal was mounted in such a way that polarization vectors of the incident soft x-ray are along the [001] axis for the horizontal polarization and along the [221] axis for the vertical polarization.

The XAS spectra can be analyzed by the VO$_6$ cluster model calculations. In the present analysis, the O 2$p$ to V 3$d$ charge transfer process is neglected since the charge transfer energies for V$^{3+}$ and V$^{2+}$ oxides are typically  around 6 eV and 8 eV, respectively \cite{Bocquet1996,Mizokawa1996}. In the present cluster model, the ligand field splitting between $e_g$ and $t_{2g}$ level (10Dq) is fixed to 1.6 eV. The Coulomb interaction between the V 3$d$ electrons are given by the Slater integrals $F^2(3d,3d)$, and $F^4(3d,3d)$ which can be translated into Racah parameters $B$ and $C$. In the present analysis, $B$ and $C$ are set to 0.117 eV and 0.438 eV \cite{Bocquet1996,Mizokawa1996}. The Coulomb interaction between the V 2$p$ core hole and the V 3$d$ electron is expressed by the Slater integrals $F^2(2p,3d)$, $G^1(2p,3d)$, and $G^3(2p,3d)$ which are fixed to 4.85 eV, 3.51 eV, and 2.00 eV, respectively (about 80\% of the atomic Hartree-Fock values) \cite{deGroot1990}. When the effect of charge transfer process is examined, the transfer integrals are parameterized by Slater-Koster parameters ($pd\sigma$) and ($pd\pi$) with ($pd\pi$)/($pd\sigma$)=-0.45. In the final states with core hole, the magnitude of transfer integrals are multiplied by 0.8 considering the contraction of 3$d$ wave functions due to core hole potential \cite{Okada1995,Mizokawa1995}. For V$^{3+}$ case, the charge transfer energy $\Delta$(multiplet averaged), V 3$d$-3$d$ Coulomb interaction $U$(multiplet averaged), and V 2$p$-3$d$ Coulomb interaction $Q$(multiplet averaged) are fixed to 6.0 eV, 5.0 eV, and 6.0 eV, respectively. Also 10Dq due to nonorthogonality is considered \cite{Mizokawa1995}.

LDA+$U$ calculations were performed using QUANTUM ESPRESSO 5.30 \cite{QE1,QE2}. We employed pseudopotentials of V.pz-spnl-kjpaw\_psl.1.0.0.UPF, O.pz-n-kjpaw\_psl.0.1.UPF, and P.pz-n-kjpaw\_psl.0.1.UPF. $U$ and $J$ are fixed to 5.0 eV and 0.5 eV, respectively. Cutoff energy was set to 30 Ry.

\section{Results and Discussion}
Figure 2 shows O 1$s$ XAS spectra taken with polarization vector along [001] direction (dashed curves) and with that along [221] direction (solid curves) at various temperatures well below the structural transition temperature at 605 K. The spectra are normalized with respect to intensities at 529 eV and 558 eV. The near edge region from 530 to 535 eV corresponds to the excitations from O 1$s$ to O 2$p$ mixed into unoccupied V 3$d$ levels. In case of LiVO$_2$ with octahedrally coordinated V$^{3+}$($t_{2g\uparrow}^2$), the transition to $t_{2g\uparrow}^3$ and those to $t_{2g\uparrow}^2e_{g\uparrow}$, $t_{2g\uparrow}^2t_{2g\downarrow}$, and  $t_{2g\uparrow}^2e_{g\downarrow}$ are located around 530 eV and 532 eV respectively \cite{Pen1997}. In the O 1$s$ XAS spectra of V$_2$OPO$_4$, the broad and weak peak at 531 eV corresponds to the transition to $t_{2g\uparrow}^3$ at V$^{3+}$. The peaks at 533 eV and 534 eV can be assigned to the transitions to $t_{2g\uparrow}^2e_{g\uparrow}$, $t_{2g\uparrow}^2t_{2g\downarrow}$, and  $t_{2g\uparrow}^2e_{g\downarrow}$ at V$^{3+}$ and the transitions from $t_{2g\uparrow}^3$ to $t_{2g\uparrow}^3e_{g\uparrow}$, $t_{2g\uparrow}^3t_{2g\downarrow}$, and  $t_{2g\uparrow}^3e_{g\downarrow}$ at V$^{2+}$. 

The peak at 531 eV is enhanced with the polarization vector along the [001] axis as shown in the difference spectra in Fig. 2(c). This polarization dependence of the 531 eV peak is more clearly seen in the bulk sensitive TFY spectra in Fig. 2(d) and indicates that the unoccupied $t_{2g}$ orbital at V$^{3+}$ is mainly mixed with O 2$p_z$ orbitals along the [001] axis.
Therefore, the unoccupied $t_{2g\uparrow}$ orbital in the V$^{3+}$($t_{2g\uparrow}^2$) site has either $yz$ or $zx$ symmetry in which the $z$ axis is slightly tilted from the [001] direction and the $x$ axis is approximately along the [201] direction for the V$^{3+}$O$_6$ octahedra in the [110] chains as indicated in Fig. 1(a). As for the other half of V$^{3+}$O$_6$ octahedra in the [1$\bar{1}$0] chains, the $z'$ axis is tilted in the other direction from the [001] direction as shown in Fig. 1(a). Although the polarization dependence of the O 1$s$ XAS spectra should be the average of the two kinds of V$^{3+}$O$_6$ octahedra, it is still roughly consistent with the unoccupied $zx$ or $yz$ ($z'x'$ or $y'z'$) orbital. Indeed, the VO$_6$ octahedron for the V$^{3+}$ site is compressed along the $z$ or $z'$ direction \cite{Pachoud2018}, and therefore, the $xy$ ($x'y'$) orbital is stabilized by the ligand field.(The V-O bond length along the $z$($z'$) direction is $\sim$ 1.98 \AA  and those along the $x$($x'$) and  $y$($y'$) directions are $\sim$ 2.09 \AA and 2.05 \AA  \cite{Pachoud2018}.)
In addition to the polarization dependence of the 531 eV peak due to the $t_{2g}$ orbital ordering, the peaks at 533 eV and 534 eV also depend on the polarization probably due to the anisotropic hybridization between V 3$d$ and O 2$p$ orbitals by the distortion of the VO$_6$ octahedra. As shown in Figs. 2(c) and (d), the temperature dependence of the O 1$s$ spectra is rather small indicating the V 3$d$ electronic state is not affected by the magnetic transition at 165 K. 

Figure 3 shows V 2$p$ XAS spectra taken with polarization vector along [001] direction (dashed curves) and with that along [221] direction (solid curves) at various temperatures well below the structural transition temperature of 605 K. The spectra are normalized with respect to intensities at 510 eV and 529 eV. The multiplet structure in the near edge region around 514 eV is derived from the V$^{3+}$ component \cite{deGroot1990}. The overall multiplet structure is rather similar to that of LiVO$_2$ \cite{Pen1997} in the surface sensitive TEY spectra. However, the intensity around 523 eV is enhanced suggesting existence of V$^{2+}$. In order to demonstrate the mixed valence of V$^{2+}$ and V$^{3+}$, calculated V 2$p$ spectra for V$^{2+}$ and V$^{3+}$ are mixed to reproduce the experimental results as shown in Fig. 4. The surface sensitive TEY spectra can be explained with $\sim$ 30\% contribution of V$^{2+}$ relative to V$^{3+}$ suggesting a mixed valence state. However, the amount of V$^{2+}$ is much smaller than that expected from the charge order in the bulk. Most probably, the V$^{3+}$ component is enhanced near the surface compared to the bulk value since the surface was exposed to the air. As for the bulk sensitive TFY spectra, it is difficult to analyze the line shape quantitatively since it is affected by the self absorption effect and the saturation effect. However, as shown in Fig. 3, the intensities around 516 eV and 523 eV (corresponding to 6 eV and 13 eV in the calculation shown in Fig. 4) are enhanced in the TFY spectra compared to the TEY spectra, indicating that the V$^{2+}$ component increases in the bulk sensitive TFY spectra compared to that in the surface sensitive TEY spectra. 

In the near edge regions around 514 eV for 2p$_{3/2}$ and around 521 eV for 2p$_{1/2}$, the V 2$p$ spectral weight is slightly enhanced with polarization vector along the [221] direction. Since the near edge regions correspond to the transitions from V 2$p$ to unoccupied V 3$d$ $t_{2g}$ at the V$^{3+}$ site, the polarization dependence in the near edge regions includes information on the V 3$d$ orbital order. In V$_2$O$_3$, Park {\it et al.} revealed that the intensity of the near edge regions is enhanced when the $a_{1g}$ orbital along the $c$ axis is unoccupied and the polarization vector is parallel to the $c$ axis \cite{Park2000}. In case of Ti$_2$O$_3$, the intensity of the near edge regions is enhanced if the $a_{1g}$ orbital along the $c$ axis is occupied and the polarization vector is perpendicular to the $c$ axis \cite{Chang2018}. Here, the $a_{1g}$ orbital is given by a linear combination of the $xy$, $yz$, and $zx$ orbitals as $(xy+yz+zx)/\sqrt{3}$. Since the $a_{1g}$ orbital does not change its sign under the rotation about the principal axis or the (111) axis, the relationship between the polarization and the intensity is rather simple.  

In the present case of V$_2$OPO$_4$, the intensity of the near edge regions is enhanced when the polarization vector is roughly along the $x$ or $y$ axis. This could indicate that the $xy$ orbital is unoccupied at the V$^{3+}$ site. However, since the $xy$ orbital changes its sign by the 90 degrees rotation about the $z$ axis, the relationship between polarization dependence and orbital occupation is less clear than for the $a_{1g}$ orbital. Figure 5 shows the calculated V 2$p$ XAS spectra for the polarizations along the $x$ and $z$ axes for the unoccupied $zx$ case and the unoccupied $xy$ case. In the unoccupied $zx$ orbital case, the near edge regions at 514 eV and 521 eV are slightly enhanced with the polarization vector along the $x$ axis, which is consistent with the experimental result. As for the unoccupied $xy$ orbital case, the near edge regions are completely suppressed with the polarization vector along the $z$ axis inconsistent with the experimental result. The comparison between the experiment and the calculation supports the orbital order with the $zx$ (or $yz$) orbital unoccupied. The occupied $xy$ orbitals provide a kind of repulsive interaction to the V$^{2+}$ and V$^{3+}$ pairs which is removed by the charge transfer between them. Therefore, the present orbital ordering is consistent with the NTE.

In the V chains running along the [110] or [1$\bar{1}$0] directions, the V$^{2+}$O$_6$ and V$^{3+}$O$_6$ octahedra share their faces. Focusing on the V chains along the [110] directions, the XAS results indicate that the V 3$d$ $t_{2g}$ $xy$ and $yz$ (or $zx$) orbitals are occupied at the V$^{3+}$ site. Here, the $x$, $y$, and $z$ axes are close to the [201], [021], and [001] directions of the monoclinic structure as shown in Fig. 1(a). (The $z$ direction is tilted by about 20 degrees from the [001] direction.) Since the superexchange pathways for the $xy$, $yz$, and $zx$ orbitals are all active in the face sharing V-O-V bond, the V$^{3+}$ and V$^{2+}$ sites should have the antiferromagnetic coupling in agreement with the neutron result \cite{Pachoud2018}. This situation is different from the ferromagnetic coupling between the face sharing V$^{3+}$ and V$^{2+}$ in BaV$_{10}$O$_{15}$ and related systems where the $a_{1g}$ orbital is unoccupied at the V$^{3+}$ site \cite{Yoshino2017,Dash2017,Dash2019}.

It is possible to analyze V 2$p$ XAS considering charge transfer from O 2$p$ to V 3$d$ if charge transfer satellites are observed at higher energy region. In the present case, the higher energy region of the V 2$p$ spectrum is occupied by the O 1$s$ spectrum, and consequently, charge transfer satellites cannot be observed. Here, we theoretically examined the effect of charge transfer process on the multiplet structure. Without the charge transfer process, the multiplet structure of V 2$p$ XAS for V$^{3+}$($d^2$) strongly depends on the ligand field splitting $10Dq$ as shown in Fig. 6. Therefore, $10Dq$ is an adjustable parameter in the calculation without charge transfer process. Since the ligand field splitting is mainly due to the covalency between the transition-metal $d$ orbitals and the ligand orbitals, $10Dq$ can be replaced by the parameters such as $\Delta$, $U$, and ($pd\sigma$) in  the calculation with charge transfer process.
In addition to the covalency effect, there exists additional $10Dq$ which can be assigned to the effect of nonorthogonality between the transition-metal $d$ orbitals and the ligand orbitals and can be obtained from the transfer and overlap integrals \cite{Mizokawa1995}. With $\Delta$=6.0 eV, $U$=5.0 eV, ($pd\sigma$)=-2.0 eV, and $10Dq$=1.0 eV due to nonorthogonality, the multiplet structure roughly resembles that obtained with $10Dq$=1.6 eV by the calculation neglecting charge transfer. This result supports the analysis neglecting charge transfer in the previous paragraphs.

For face sharing octahedra in the chains, it is often convenient to use $a_{1g}$ and $e_g^{\pi}$ trigonal orbitals. Their forms are given by $(xy-e^{2\pi ni/3}zx-e^{-2\pi ni/3}yz)/\sqrt{3}$, with $n=0$ for the $a_{1g}$ state and $n=\pm 1$ for two $e_{g\pm}^{\pi}$ orbitals (The $a_{1g}$ orbital points to the (1,1,-1) direction of the VO$_6$ octahedron for the case shown in Fig. 1) \cite{Khomskii2014}. One can show that the electron hopping along chains are diagonal for these orbitals \cite{Kugel2015}, which immediately gives the dominant antiferromagnetic coupling along the chains if the same orbitals are occupied at each site. However, the use of these orbitals is not convenient to treat interchain exchange. Besides that, as mentioned above, in V$_2$OPO$_4$ there exists strong distortion of VO$_6$ octahedra along the local $z$ (or $z'$) directions. Therefore, it is more natural to use the local $xy$, $zx$ and $yz$ orbitals that we use in this paper.
The $xy$ and $yz$ (or $zx$) orbitals are given by $(a_{1g}+e_{g+}^{\pi}+e_{g-}^{\pi})/\sqrt{3}$ and $-(a_{1g}+e^{2\pi i/3}e_{g+}^{\pi}+e^{-2\pi i/3}e_{g-}^{\pi})/\sqrt{3}$ [or $-(a_{1g}+e^{-2\pi i/3}e_{g+}^{\pi}+e^{2\pi i/3}e_{g-}^{\pi})/\sqrt{3}$].  Therefore, in the electronic configuration with $xy$ and $yz$ occupied, the $a_{1g}$, $e_{g+}^{\pi}$, and $e_{g-}^{\pi}$ orbitals are equally occupied. Since the $a_{1g}$, $e_{g+}^{\pi}$, and $e_{g-}^{\pi}$ electrons at the V$^{2+}$ site can be transferred to the $a_{1g}$, $e_{g+}^{\pi}$, and $e_{g-}^{\pi}$ orbitals at the V$^{3+}$ site, the superexchange interaction is antiferromagnetic.

In case of the V chains along the [1$\bar{1}$0] directions, the $z'$ axis is close to the [001] direction and the polarization vector in the experiment. It can be assumed that the V 3$d$ $t_{2g}$ $x'y'$ and $y'z'$ (or $z'x'$) orbitals are occupied at the V$^{3+}$ site in the [1$\bar{1}$0] chains.
As for the superexchange pathways between the V$^{2+}$ site in the [110] chain and the V$^{3+}$ site in the [1$\bar{1}$0] chain (the V-O-V bond angle is about 124 degrees), all the $t_{2g}$ electrons of the V$^{2+}$ site can be transferred to the $x'y'$ and $y'z'$ (or $z'x'$) orbitals at the V$^{3+}$ site which are already occupied by electrons. Therefore, the superexchange interaction is expected to be antiferromagnetic which is also consistent with the neutron result \cite{Pachoud2018}. 

The superexchange pathways between the two V$^{3+}$ sites in the [110] and [1$\bar{1}$0] chains are mainly given by those between $yz$ and $y'z'$ orbitals (also $zx$ and $z'x'$ orbitals) since the V-O-V bond angle is about 132 degrees [see Fig. 1(a)]. If the $zx$ orbital is unoccupied in the chains running along [110] and the $y'z'$ orbital is unoccupied in the chains running along [1$\bar{1}$0], the two V$^{3+}$ sites along the (001) direction should be ferromagnetic due to the Kugel-Khomskii mechanism \cite{Kugel1972}. On the other hand, if the $zx$ orbital is unoccupied in the chains running along [110] and the $z'x'$ orbital is unoccupied in the chains running along [1$\bar{1}$0], the superexchange interaction would be antiferromagnetic. The ferromagnetic V$^{3+}$-V$^{3+}$ coupling is consistent with the neutron result \cite{Pachoud2018}. Since the charge transfer from the V$^{2+}$ site to the V$^{3+}$ site costs smaller energy than the other charge transfer processes, the antiferromagnetic couplings between the V$^{2+}$ and V$^{3+}$ sites (both intrachain and interchain) are essential to determine the spin arrangement.

Figures 7(a) and (b) show V 3$d$ and O 2$p$ partial density of states of the ferrimagnetic solution with $U$=5.0 eV and $J$=0.5 eV. V 3$d$ electrons of the V$^{2+}$ (V$^{3+}$) sites mainly contribute to the majority (minority) spin density of states. In Fig. 7(a), the occupied band around -2 eV (minority spin, V$^{3+}$) is dominated by the $XY$ component. The unoccupied band around 1 eV (minority spin, V$^{3+}$) is a mixture of $ZX$ and $YZ$ components. Here, the $X$ and $Y$ axes are along the a and b axes, and the $Z$ axis in the ac plane. The calculated orbital occupation of the V$^{3+}$ site is consistent with the XAS result. It is not obvious why the $ZX$ component is higher than the $YZ$ component in the unoccupied band. We speculate that the global monoclinic structure plays a role rather than the distortion of VO$_6$ octahedron. The O 2$p$ partial density of states is consistent with the lineshape of O 1$s$ XAS. The lowest energy unoccupied band around 1 eV corresponds to the transition to $t_{2g\uparrow}^3$ at V$^{3+}$.
The unoccupied bands ranging from 2 to 4 eV correspond to the transitions to $t_{2g\uparrow}^2e_{g\uparrow}$, $t_{2g\uparrow}^2t_{2g\downarrow}$, and  $t_{2g\uparrow}^2e_{g\downarrow}$ at V$^{3+}$ and the transitions from $t_{2g\uparrow}^3$ to $t_{2g\uparrow}^3e_{g\uparrow}$, $t_{2g\uparrow}^3t_{2g\downarrow}$, and  $t_{2g\uparrow}^3e_{g\downarrow}$ at V$^{2+}$.
In the lowest energy unoccupied band at $\sim$ 1eV, the amount of O 2$p$ $Y$ is much smaller than those of $X$ and $Z$ components. This is roughly consistent with the polarization dependence of O 1$s$ XAS although the $Z$ axis is tilted by  $\sim$ 30$^{\circ}$ from the c axis. The consistency between the VO$_6$ cluster model and the band structure supports the interpretation of the O 1$s$ XAS result \cite{Lafuerza2011}.

\section{Conclusion}
In summary, the polarization dependence in near edge regions of O 1$s$ and V 2$p$ XAS spectra indicates that the $xy$ and $zx$ (or $yz$) orbitals are occupied in the V$^{3+}$ site. The V 3$d$ orbital order is consistent with the antiferromagnetic coupling between the face sharing V$^{2+}$ and V$^{3+}$ sites in the chains and the antiferromagnetic coupling between the corner sharing V$^{2+}$ and V$^{3+}$ sites between the chains. In future, the relationship between the orbital order and the negative thermal expansion should be investigated by theoretical calculations and experiments for higher temperature.

\section*{Acknowledgements}
The authors would like to thank Dr. Feizhou He for the technical support at the REIXS beamline of the Canadian Light Source. All of the measurements reported in this paper were performed at the Canadian Light Source, a national research facility of the University of Saskatchewan, which is supported by the Canada Foundation for Innovation (CFI), the Natural Sciences and Engineering Research Council (NSERC), the National Research Council (NRC), the Canadian Institutes of Health Research (CIHR), the Government of Saskatchewan, and the University of Saskatchewan. The work at Waseda was partially supported by CREST-JST (Grant No. JPMJCR15Q2) and KAKENHI from JSPS (Grants No.19H01853 and No.19H00659). The work at Edinburgh was supported by EPSRC and ERC. The work of D.I. Khomskii was funded by the Deutsche Forschungsgemeinschaft (DFG, German Research Foundation) - Project number 277146847 - CRC 1238.

\clearpage

\begin{figure}
\includegraphics[width=10cm]{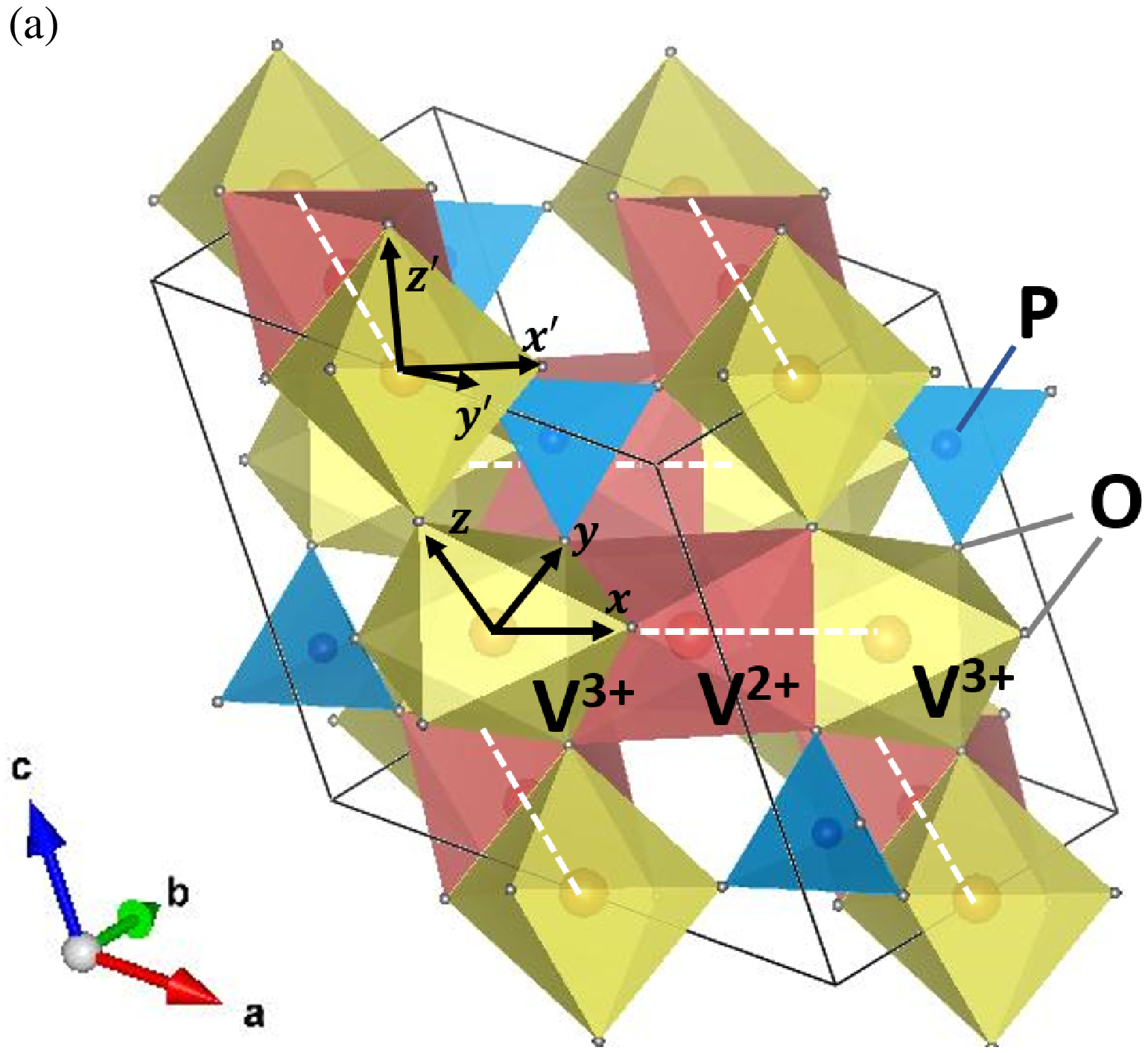}\\
\includegraphics[width=5cm]{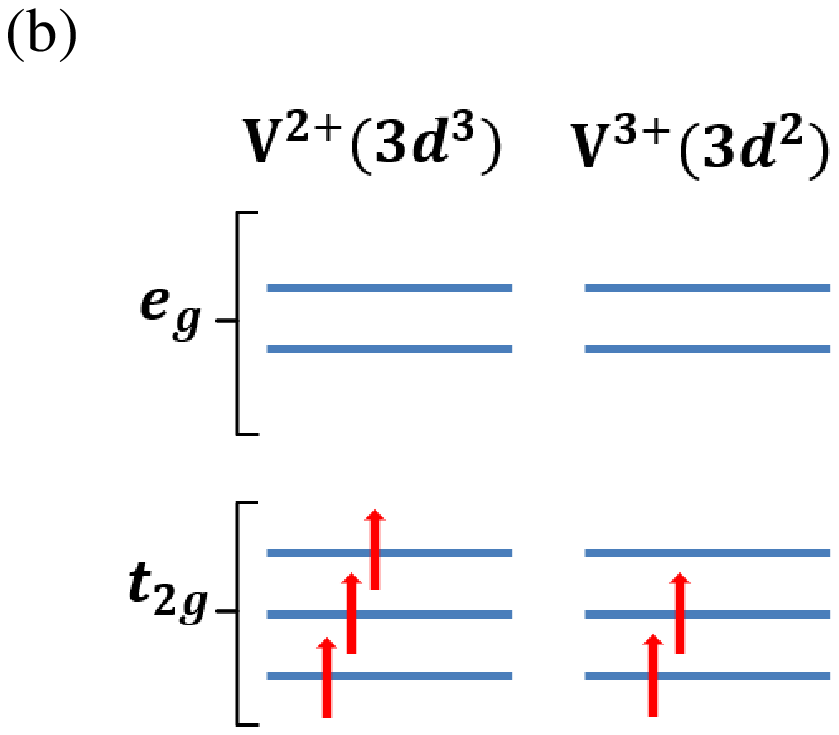}
\includegraphics[width=5cm]{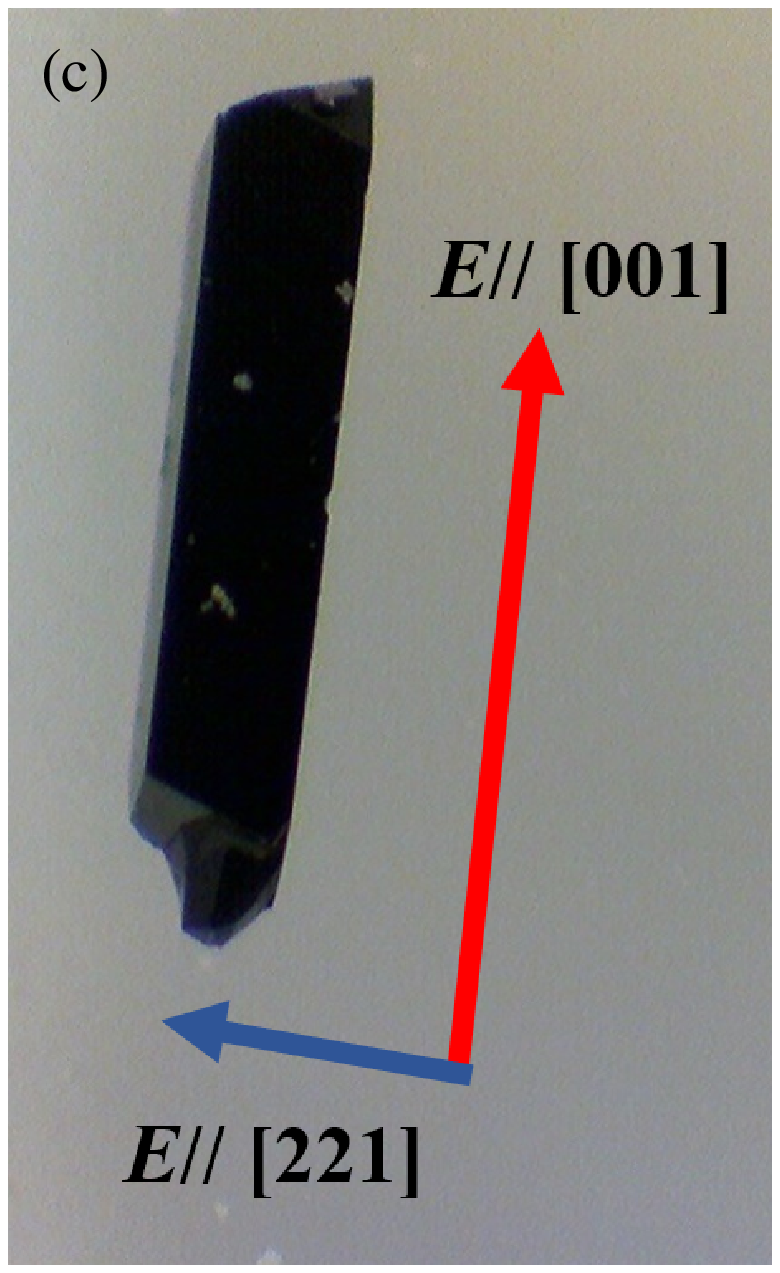}
\caption{
(a) Crystal structure of V$_2$OPO$_4$ created by VESTA \cite{VESTA}. The dashed lines indicate the face sharing VO$_6$ chains along [110] or [1$\bar{1}$0] directions.
The $x$, $y$, and $z$ axes ($x'$, $y'$, and $z'$ axes) are along the V-O bonds of the V$^{3+}$ site in the [110] chain ([1$\bar{1}$0] chain).
(b) Electronic configurations for V$^{2+}$ and V$^{3+}$.
(c) Photograph of the V$_2$OPO$_4$ crystal and the directions of polarization vector.
}
\end{figure}
\clearpage

\begin{figure}
\includegraphics[width=8cm]{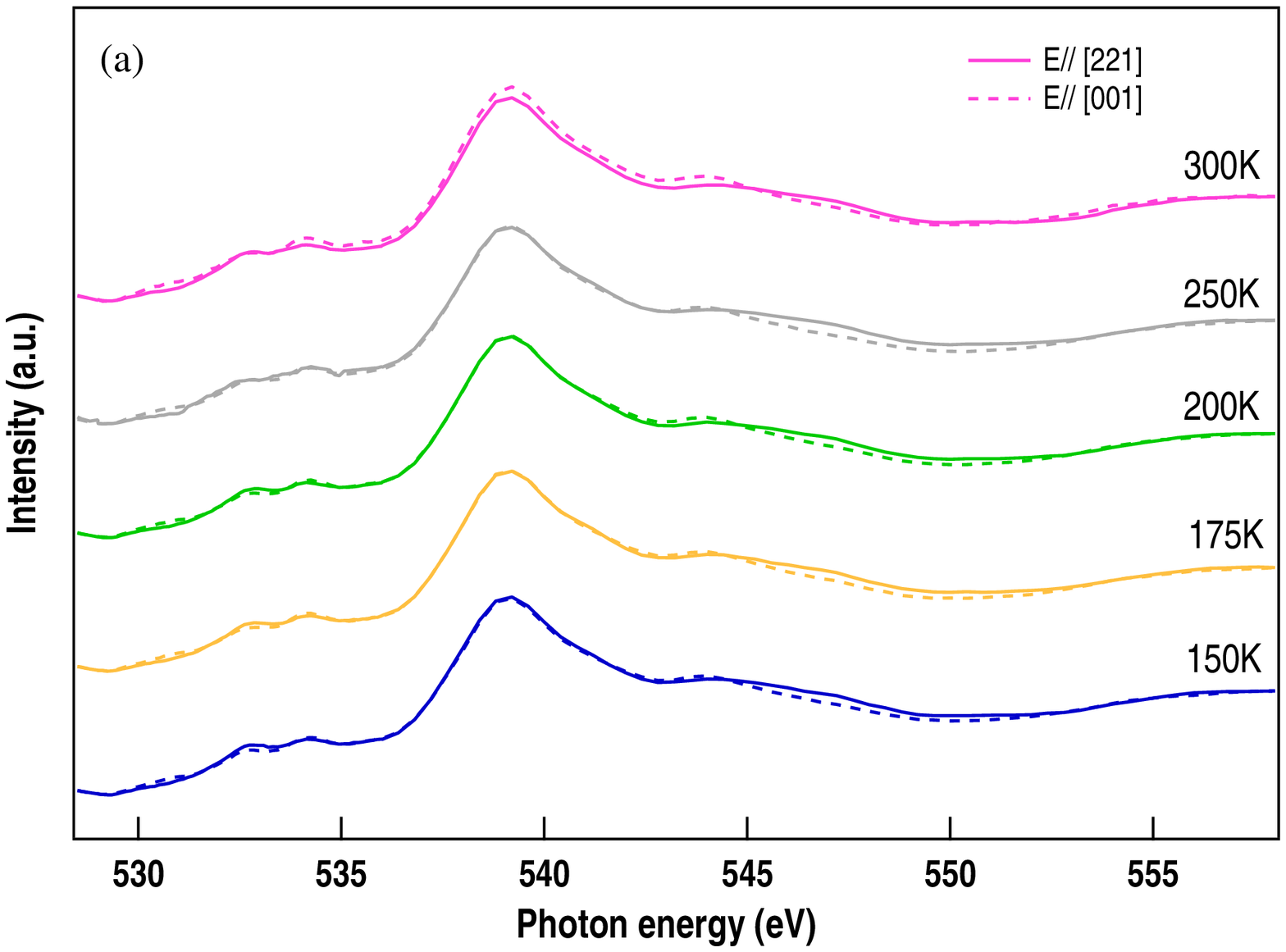}
\includegraphics[width=8cm]{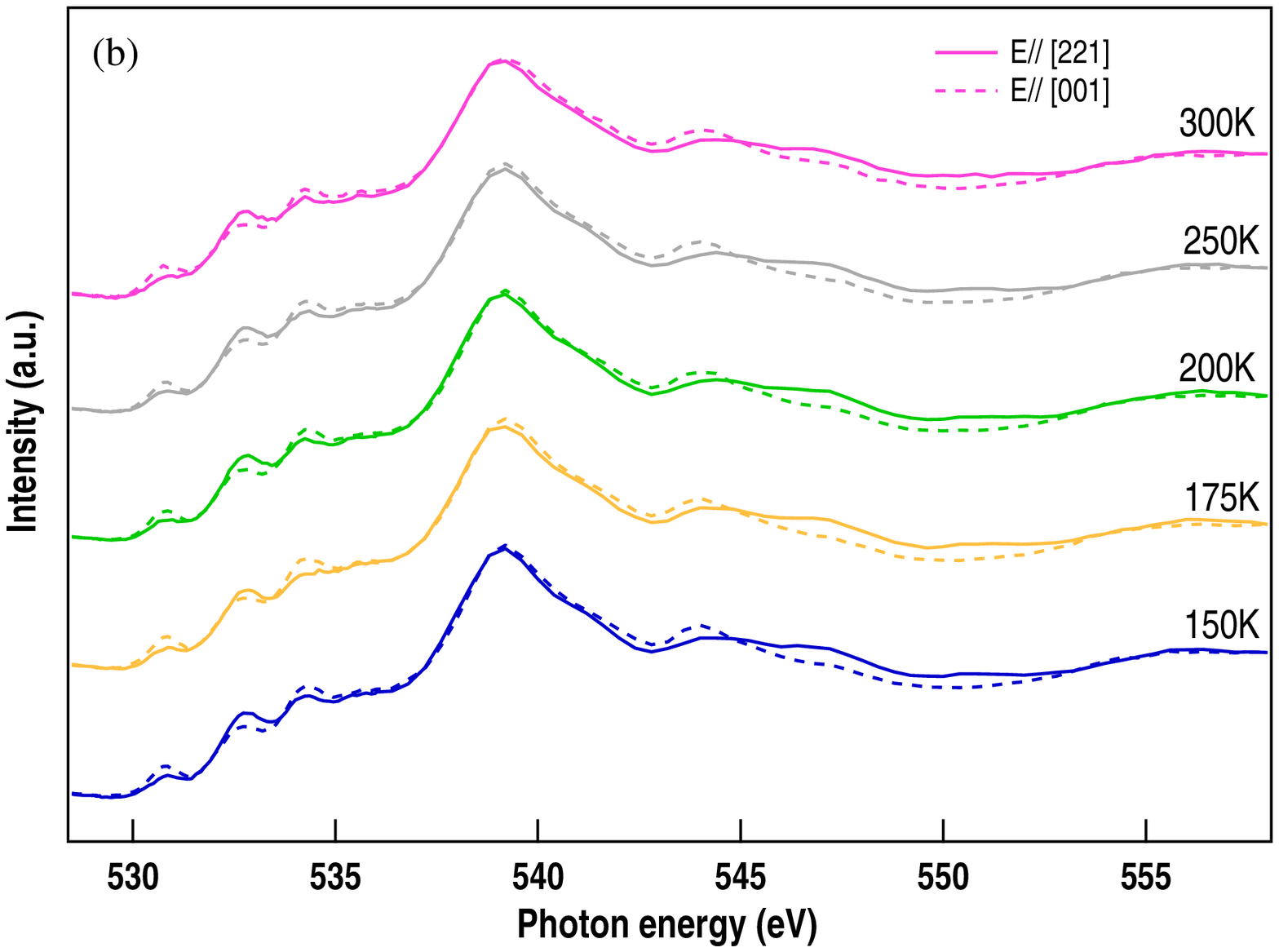}
\includegraphics[width=8cm]{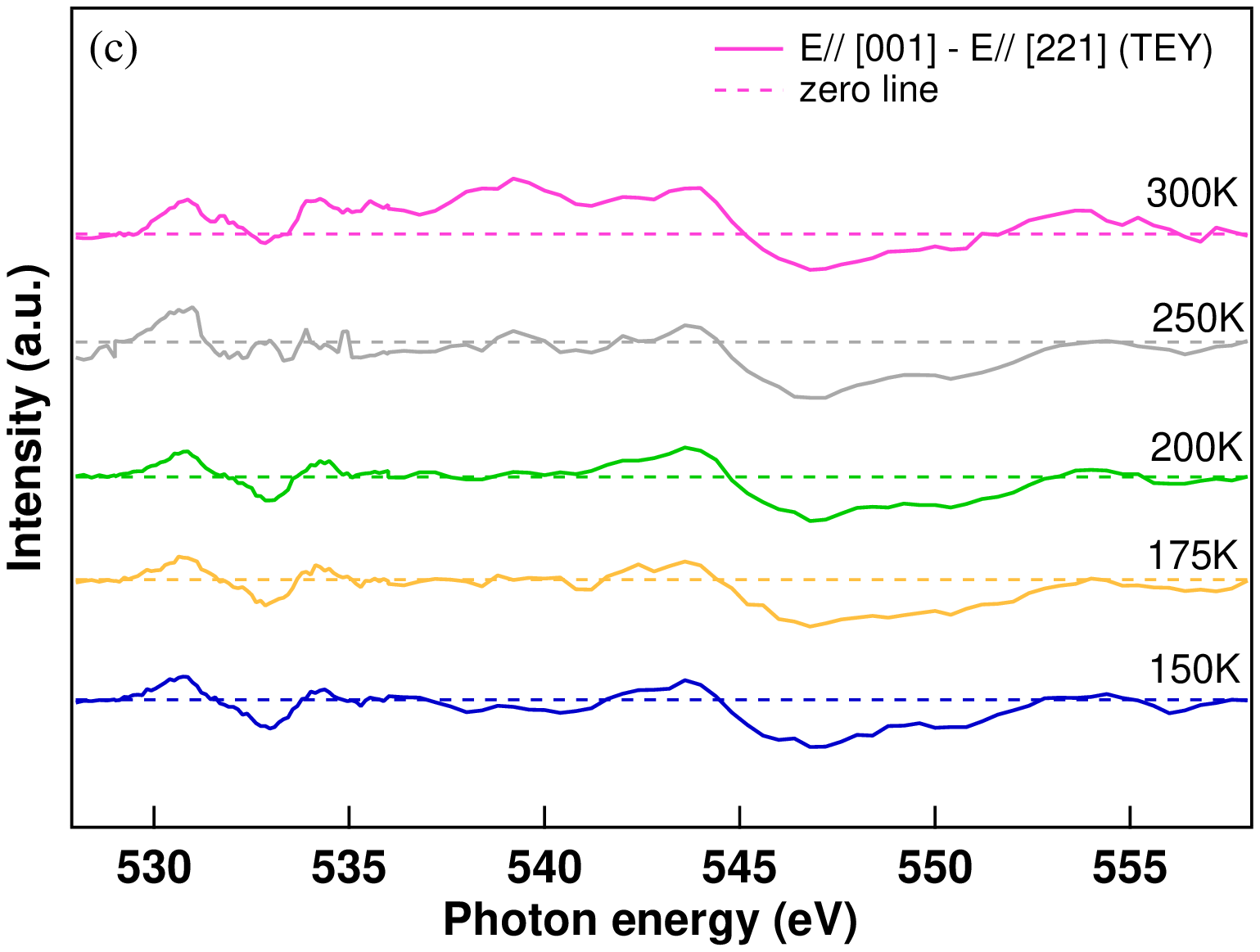}
\includegraphics[width=8cm]{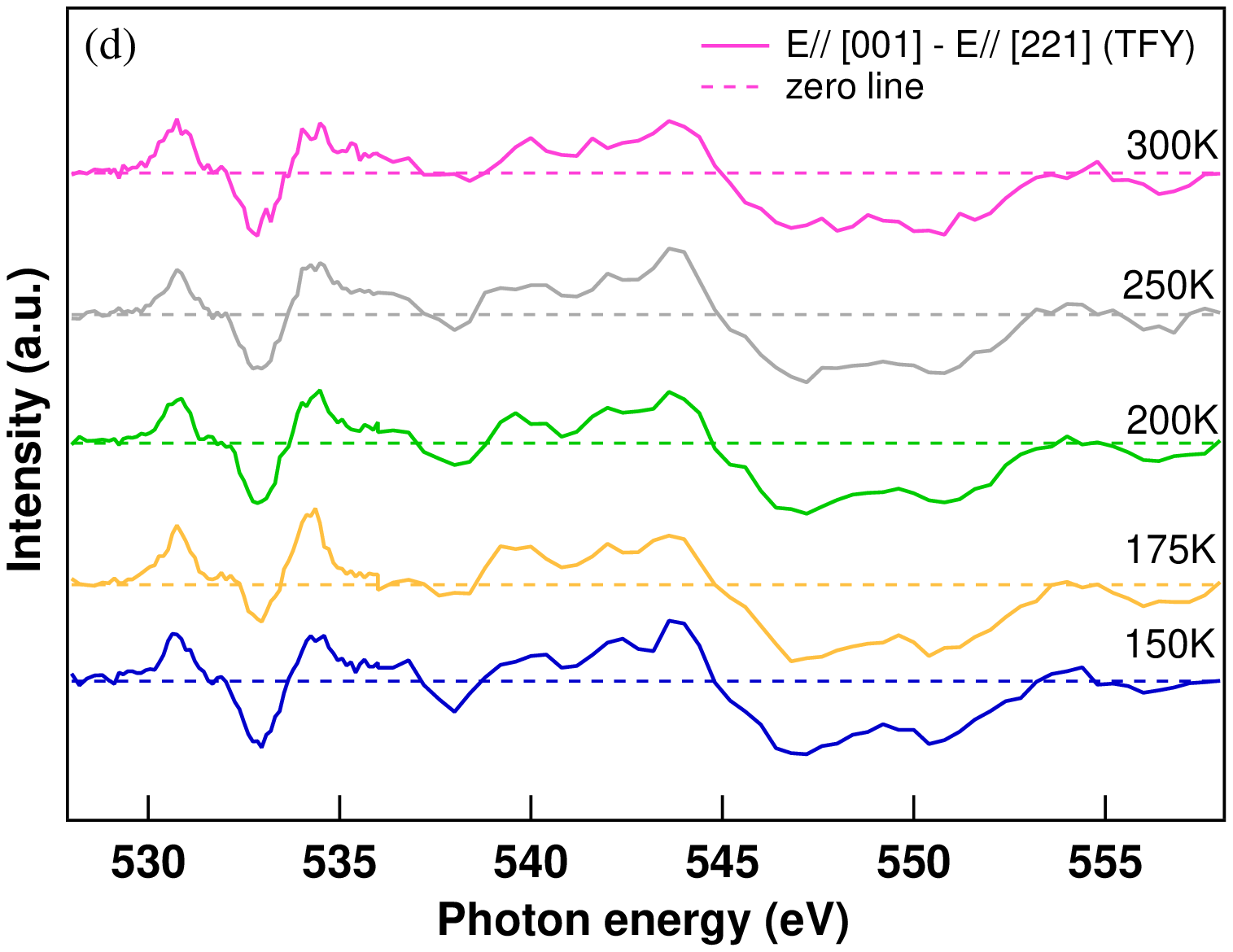}
\caption{
(a) Temperature and polarization dependence of O $1s$ XAS with TEY mode.
(b) Temperature and polarization dependence of O $1s$ XAS with TFY mode.
(c) Temperature dependence of linear dichroism in O $1s$ XAS with TEY mode.
(d) Temperature dependence of linear dichroism in O $1s$ XAS with TFY mode.
}
\end{figure}
\clearpage

\begin{figure}
\includegraphics[width=10cm]{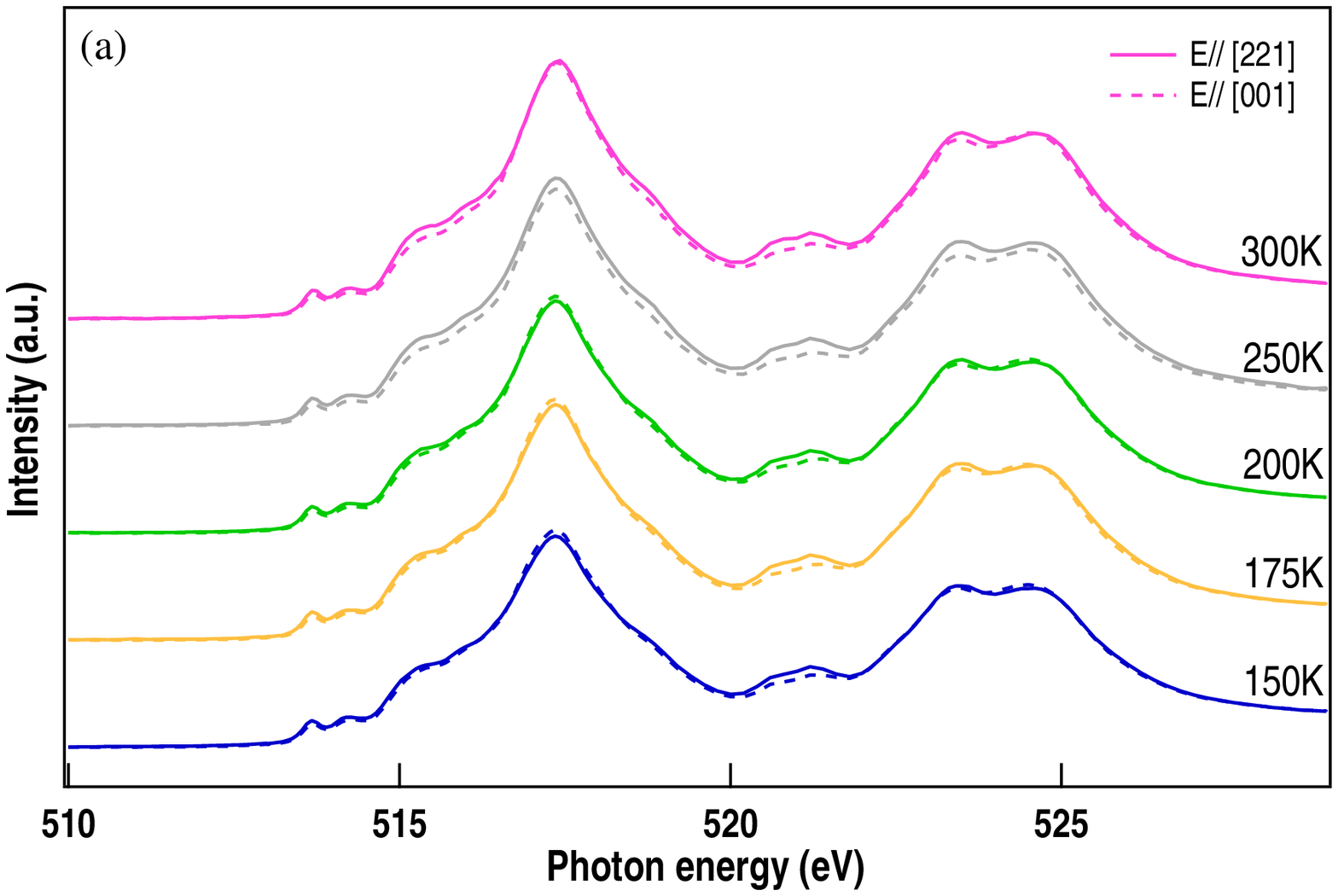}
\includegraphics[width=10cm]{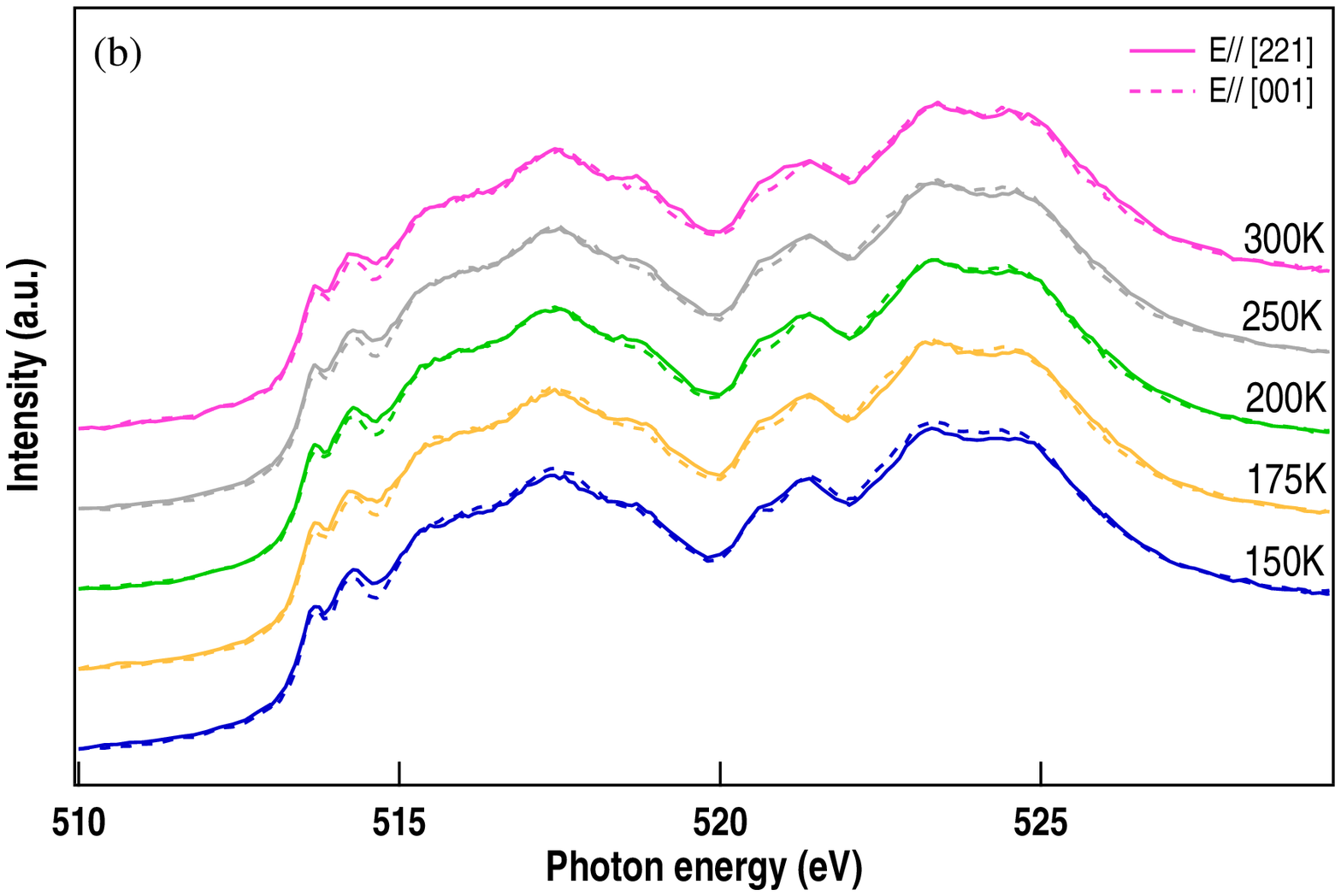}
\caption{
(a) Temperature and polarization dependence of V 2$p$ XAS with TEY mode.
(b) Temperature and polarization dependence of V 2$p$ XAS with TFY mode.
}
\end{figure}
\clearpage

\begin{figure}
\includegraphics[width=10cm]{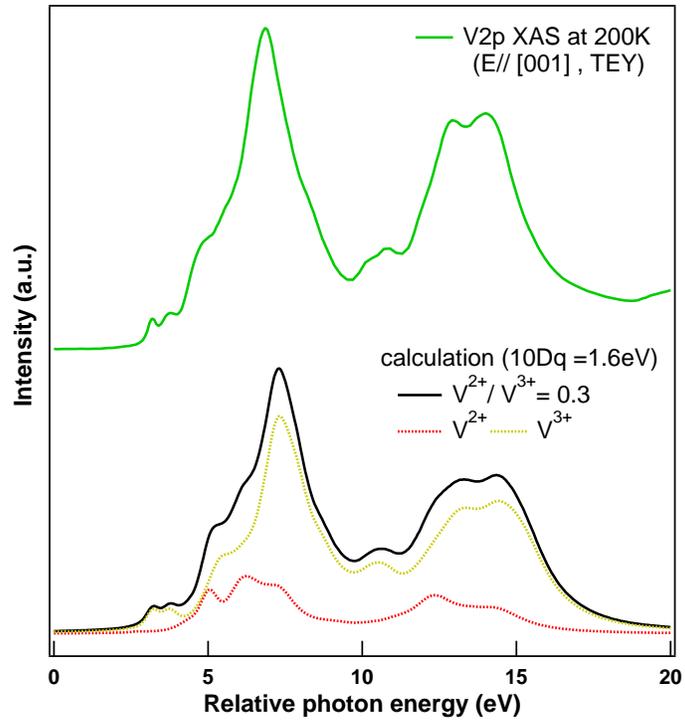}
\caption{
Cluster model calculation of V 2$p$ XAS for a mixed valence state of V$^{2+}$ : V$^{3+}$ = 3 : 10. 
The calculated spectrum is compared with the experimental result taken at 200 K with TEY mode.
}
\end{figure}
\clearpage

\begin{figure}
\includegraphics[width=10cm]{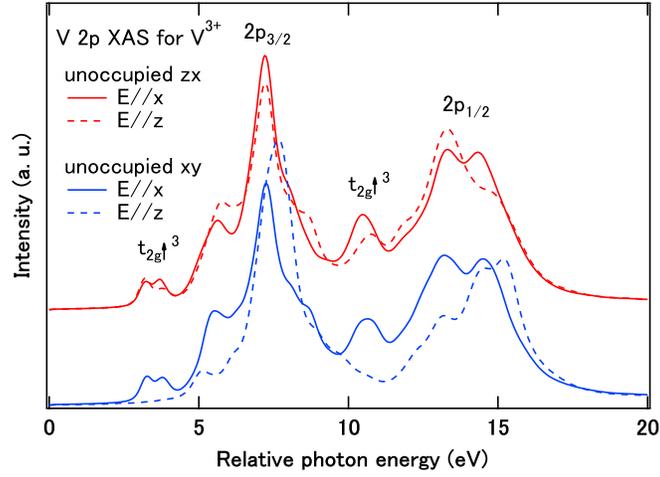}
\caption{
Polarization dependence of the calculated V 2$p$ XAS spectra for V$^{3+}$ with different orbital orders. 
For the orbital order with V 3$d$ $yz$ unoccupied, polarization dependence of the $t_{2g\uparrow}^3$ final state is moderate. 
On the other hand, the orbital order with V 3$d$ $xy$ unoccupied is expected to have strong polarization dependence.
}
\end{figure}
\clearpage

\begin{figure}
\includegraphics[width=10cm]{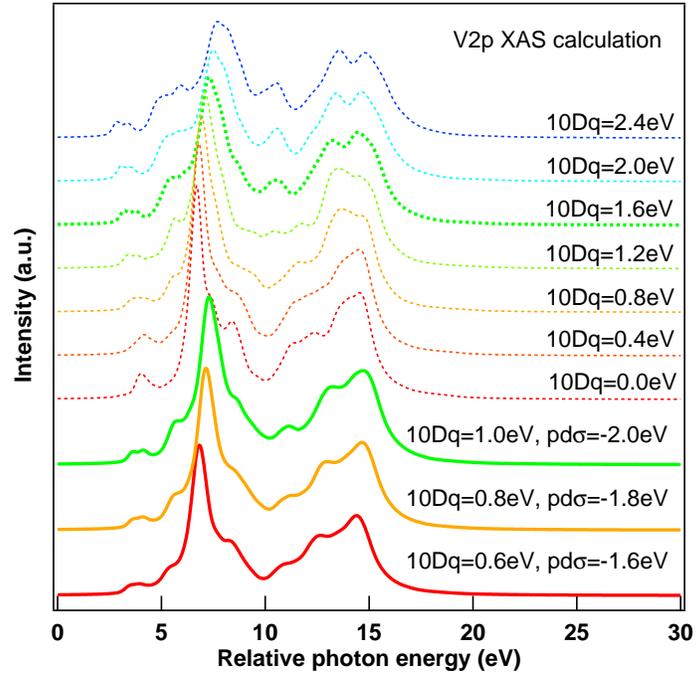}
\caption{
Effect of charge transfer process on V 2$p$ XAS calculation for V$^{3+}$.
The dotted (solid) curves indicate the calculations without (with) the charge transfer process.
}
\end{figure}
\clearpage

\begin{figure}
\includegraphics[width=10cm]{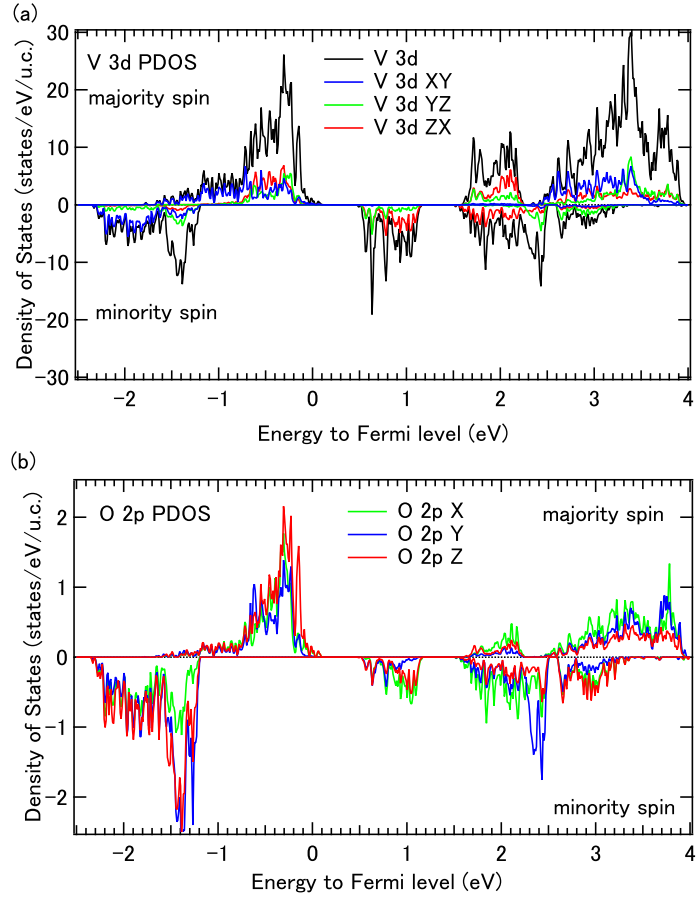}
\caption{
(a)V 3$d$ partial density of states and (b) O 2$p$ partial density of states obtained by LDA+$U$ with $U$=5.0 eV and $J$=0.5 eV.
}
\end{figure}

\end{document}